\documentclass[prd,twocolumn,showpacs,amsmath,amssymb]{revtex4}

\usepackage{amssymb}
\usepackage{mathrsfs}
\usepackage{txfonts}

\usepackage{graphicx}
\usepackage{dcolumn}
\usepackage{bm}

\begin{document}

\title{Perturbation spectra in the warm $k$-inflation}

\author{Zhi-Peng Peng\textsuperscript{1} }
\email{zhipeng@mail.bnu.edu.cn}
\author{Jia-Ning Yu\textsuperscript{2}}
\email{yujn@mail.bnu.edu.cn}
\author{Xiao-Min Zhang\textsuperscript{3}}
\email{zhangxm@mail.bnu.edu.cn}
\author{Jian-Yang Zhu\textsuperscript{2}}
\thanks{Corresponding author}
\email{zhujy@bnu.edu.cn}

\affiliation{
\textsuperscript{1}Gravitational Wave and Cosmology Laboratory, Department of Astronomy, Beijing Normal University, Beijing 100875, China \\
\textsuperscript{2}Department of Physics, Beijing Normal University, Beijing 100875, China \\
\textsuperscript{3}School of Science, Qingdao University of Technology, Qingdao 266033, China
}

\date{\today}

\begin{abstract}
We investigate the cosmological perturbation theory and calculate the perturbation spectra in the warm $k$-inflation. Unlike the scalar perturbations of $k$-inflation in the cold inflationary scenario, entropy perturbations should be considered in our model. We study how the entropy perturbations affect the evolution of curvature perturbation $\mathcal{R}$ on the comoving hypersurfaces and find the dissipation coefficient $\Gamma$ plays a central role in the entropy perturbations and the evolution equation of $\mathcal{R}$. In the weak dissipation condition, entropy perturbations on the large scale can be neglected such that the primordial spectrum of cosmological perturbations is due only to adiabatic perturbations. Like general warm inflationary models, density fluctuations also mainly originated from the thermal fluctuations of the inflaton field rather than the vacuum fluctuations in our model. Finally, we calculate the perturbation spectra and the tensor-to-scalar ratio, then find the tensor-to-scalar ratio is smaller than that in the $k$-inflation or the general potential-driven cold inflationary models.
\end{abstract}

\maketitle

\section{\label{sec1}Introduction}
Inflation is the most successful theory that we have available for explaining the large scale structure of the Universe. In the previous studies, inflationary dynamics were realized in two different methods. In the initial picture, termed "standard inflation" or "cold inflation", the Universe rapidly supercools during an inflation stage and subsequently a reheating stage is invoked to end inflation and fill the Universe with radiation \cite{Guth1981,Linde1982,Albrecht1982,Hawking1982}. In the other picture, termed "warm inflation", the radiation is produced throughout inflation so that its energy density is kept nearly constant in this phase. This is accomplished by introducing a dissipation term $\Gamma$ in the equation of motion for the inflaton field, considering a continuous energy transfer from its decay. As a result, no reheating is necessary, since we expect that enough radiation is produced to provide a smooth transition to the radiation era at the end of inflation \cite{Berera1995,Berera1996,Berera1997}.

In cold inflation, it is generally considered that inflation produced seeds that give rise to the large-scale structure and the observed little anisotropy of the cosmological microwave background (CMB) through the vacuum fluctuations of inflaton field \cite{Ellis1989,Liddle2000,Liddle2001,Weinberg}. However, the seeds are no longer generated via quantum fluctuations in warm inflation but are dominated by the larger thermal fluctuations \cite{Berera1995,Moss1985,Berera2000}. These have their origin in the hot radiation and influence the inflaton through a thermal noise term $\xi(x,t)$ in the equation of motion.

In general research concerned with inflation, the potential energy of the inflaton field is the dominating energy and drives the inflation during the inflationary epoch in the cold or warm inflation. However, a novel model called ``$k$-inflation'' provided by Mukhanov \cite{Mukhanov1999} only has a general class of nonstandard (i.e., nonquadratic) kinetic energy terms for a scalar field, which is a kind of kinetically driven inflationary model. Recently, we have extended $k$-inflation in the cold inflationary scenario to a possible warm inflationary scenario, so called warm $k$-inflation \cite{Peng2016}. And the theory of cosmology perturbations in the $k$-inflation has been extended from that of the potential-driven model by Mukhanov \cite{Mukhanov}. However, it is very different from that in warm inflation, since during the warm inflationary phase, the constitutive components of the Universe are mainly two fields including the inflaton field and radiation, not only an inflaton field. In the two-field models like warm inflationary models, we are obliged to take into account entropy perturbations \cite{Bardeen1980,Mukhanov1992,Gordon2000,Hwang2000,Taylor2000,Malik2003,Hall2004,Malik2005}.
The existence of entropy perturbations may provide a nontrivial evolution of the comoving curvature perturbation $\mathcal{R}$, because this quantity is no longer frozen for large-scale modes as in single-field inflationary models \cite{Oliveria2001}.

In this paper, we introduce the theory of cosmology perturbations and calculate the perturbation spectra  in the warm $k$-inflation. First, we give the dynamical equations of warm $k$-inflation in the unperturbed Friedman-Robertson-Walker (FRW) background and the results of stability analysis of the system that have been discussed in Ref.\cite{Peng2016}. Second, we study scalar perturbations of the FRW metric in comparison with general potential-driven inflationary models and $k$-inflation. Like the cosmology perturbation theory in the two-field inflation, entropy perturbations must be considered in the warm $k$-inflation. We find the entropy perturbations can be divided into two parts that are caused by the metric perturbation and the thermal dissipation, respectively. The dissipation term $\Gamma$ plays an important role in the entropy perturbations. Since if $\Gamma$ is of absence, the radiation will redshift away during inflation. Then, the overall scenario is reduced to a single field, in which entropy perturbations are neglected on the large scale \cite{Sri2009}. When it is present, the radiation will be continuously produced rather than washed away. Then the comoving curvature perturbation $\mathcal{R}$ will be affected by the entropy perturbations and no longer adiabatic. However, the source term in the evolution equation of $\mathcal{R}$ can be neglected in the weak dissipation condition ($r\ll1$), which is still allowed in warm inflation, the evolution equation of $\mathcal{R}$ on the large scale will be approximately written as $\dot{\mathcal{R}}\simeq 0$ meaning that the entropy perturbations are no longer important and the primordial spectrum of the perturbations is due only to adiabatic perturbations. Third, we calculate the thermal fluctuations of the inflaton field in comparison with the vacuum fluctuations; then, we find density fluctuations also mainly originating from the thermal fluctuations of the inflaton field, which is consistent with the fundamental characteristic of cosmology perturbations in warm inflation \cite{Berera1995}. Fourth, the perturbation spectra are worked out to compare to the previous theoretical results and the observable results. And we find the tensor-to-scalar ratio is smaller than that in the $k$-inflation or the general potential-driven cold inflationary models. Finally, we show the validity of the results and the consistency of the weak dissipation assumption through a specific case.

The paper is organized as follows. In Sec.\ref{sec2}, basic dynamics of the homogenous inflaton field is briefly introduced  in the warm $k$-inflation. Then in Sec.\ref{sec3}, we will investigate the scalar perturbations of metric which are very different from those in the usual potential-driven cold inflationary models. Then, the thermal fluctuations of the inflaton field are studied in Sec.\ref{sec4}, and the spectra of scalar and tensor perturbations are discussed in Secs.\ref{sec5} and \ref{sec6}. Finally, we draw the conclusions in Sec.\ref{sec7}.

\section{\label{sec2}dynamics of the warm $k$-inflation}
We consider the most general local action for a scalar field coupled to radiation and Einstein gravity,
\begin{equation}\label{action}
  S=\int d^4x \sqrt{-g}  \left[ -\frac{R}{16\pi G}+\mathcal{L}(X,\phi)+\mathcal{L}_\gamma+\mathcal{L}_{int}\right],
\end{equation}
where $g$ is the determinant of the metric, $\mathcal{L}(X,\phi)$ is the Lagrangian density of the inflaton field, $\mathcal{L}_\gamma$ is the Lagrangian density of the radiation field, and $\mathcal{L}_{int}$ describes the interaction between the inflaton field and other fields. In this paper, we will use the signature $(+,-,-,-)$. The Lagrangian density of the inflaton field takes the same simple form as that in Refs.\cite{Mukhanov1999,Peng2016},
\begin{equation}\label{field}
  \mathcal{L}(X,\phi)=K(\phi)X+X^2,
\end{equation}
where $X=(1/2) g^{\mu\nu}\partial_{\mu}\phi\partial_{\nu}\phi$, and $K(\phi)$ is a function of the inflaton field $\phi$, called `` kinetic function ''.
The inflaton field in a spatially flat FRW Universe is described by an effective fluid with the energy-momentum tensor
\begin{equation}\label{EM1}
   T^{(\phi)}_{\mu\nu}=(\rho_{\phi}+p_{\phi})u_{\mu}u_{\nu}-p_{\phi}g_{\mu\nu},
\end{equation}
where $\rho_{\phi}$, $p_{\phi}$ and $u_{\mu}$ are the energy density, pressure, and 4-velocity of the inflaton field. Indeed, varying the Lagrangian density of the inflaton field with respect to the metric, we can obtain the energy-momentum of the inflaton field,
\begin{equation}\label{EM2}
  T^{(\phi)}_{\mu\nu}=\frac{2}{\sqrt{-g}}\frac{\delta (\sqrt{-g}\mathcal{L}(X,\phi))}{\delta g^{\mu\nu}}=\mathcal{L}_{X}\partial_{\mu}\phi\partial_{\nu}\phi-g_{\mu\nu}\mathcal{L}(X,\phi),
\end{equation}
where $\mathcal{L}_{X}$ denotes a partial derivative of the Lagrangian $\mathcal{L}(X,\phi)$ with respect to $X$. Comparing Eq.(\ref{EM2}) to Eq.(\ref{EM1}), we find
\begin{eqnarray}
  \rho_{\phi}&=&K(\phi)X+3X^2, \label{density}\\
    p_{\phi}&=&K(\phi)X+X^2, \label{pressure}\\
    u_{\mu}&=&\sigma\frac{\partial_{\mu}\phi}{\sqrt{2X}},
\end{eqnarray}
where $\sigma$ denotes the sign of $\dot{\phi}$ (when $\dot{\phi}>0$, $\sigma=1$; otherwise, $\sigma=-1$ ).

The dynamics of the FRW cosmological model in the warm $k$-inflation \cite{Peng2016} is described by the equations
\begin{equation}\label{Friedmann}
  H^2=\frac{1}{3M_{p}^2}(\rho_{\phi}+\rho_{\gamma}),
\end{equation}
\begin{equation}\label{phicon}
  \dot{\rho}_{\phi}+3H(\rho_{\phi}+p_{\phi})=-\Gamma\dot{\phi}^2,
\end{equation}
\begin{equation}\label{radiation}
  \dot{\rho}_{\gamma}+4H\rho_{\gamma}=\Gamma \dot{\phi}^2,
\end{equation}
where $M_{p}^2\equiv(8\pi G)^{-1}$, $\rho_{\gamma}$ is the energy density of radiation and $\Gamma$ is the dissipative coefficient in warm inflation. The dissipative coefficient phenomenologically describes the decay of the inflaton field $\phi$ into radiation via the interaction Lagrangian $\mathcal{L}_{int}$ during the inflationary phase.
As usual, we consider a homogeneous background scalar field $X=\frac{1}{2}\dot{\phi}^2$; then, the equation of motion of inflaton field can be obtained by substituting Eqs.(\ref{density}) and (\ref{pressure}) into Eq.(\ref{phicon}),
\begin{equation}\label{EOM}
  (3\dot{\phi}^2+K)\ddot{\phi}+3H(\dot{\phi}^2+K+r)\dot{\phi}+\frac{1}{2}K_{\phi}\dot{\phi}^2=0,
\end{equation}
where $K_{\phi}$ is a derivative of kinetic function $K(\phi)$ with respect to the inflaton field and the dissipative rate is defined as $r \equiv\frac{\Gamma}{3H}$.

Since the exact Eqs. (\ref{Friedmann}), (\ref{radiation}), and (\ref{EOM}) are difficult to solve, a slow-roll approximation is often applied. And we have performed a stability analysis to obtain the slow-roll conditions for the systems that remain close to the quasiexponential inflationary attractor within many Hubble times \cite{Peng2016}. These slow-roll conditions are
\begin{eqnarray}\label{slcon}
  \epsilon\ll1, \vert \eta \vert \ll\frac{\mathcal{L}_{X}}{(\mathcal{L}_{X}+r)c_{s}^2},\vert b \vert \ll1,\ \vert c \vert <4, \frac{rc_{s}^2}{\mathcal{L}_{X}}\ll1-2c_{s}^2
\end{eqnarray}
where the parameters $\epsilon=\frac{K_{\phi}\dot{\phi}}{HK}$, $\eta=\frac{K_{\phi\phi}\dot{\phi}}{HK_{\phi}}$, $b=\frac{\Gamma_{\phi}\dot{\phi}}{H\Gamma}$,  $c=\frac{T\Gamma_{T}}{\Gamma}$ and $c_{s}^2=\frac{\dot{\phi}^2+K}{3\dot{\phi}^2+K}\simeq \frac{\dot{p}}{\dot{\rho}}<1 $. Among the definition of the parameters, the dot denotes the derivative of quantities with respect to time, and the subscripts denote the partial derivative of the quantities with respect to the inflaton field or temperature. In addition, quasiexponential warm $k$-inflation ensures the term $\dot{\phi}^2+K+r$ is a small positive quantity and $\dot{\phi}^2$ has the same order of magnitude as $|K(\phi)|$ \cite{Peng2016}.

During the inflationary era, it is well inside slow-roll period, the energy density associated with the inflaton field is the order of $\frac{1}{4}|K|\dot{\phi}^2$, i.e. $\rho_{\phi}\sim\frac{1}{4}|K|\dot{\phi}^2$, and dominates over the energy density associated with the radiation field, i.e. $\rho_{\phi}>\rho_{\gamma}$. Thus, the Friedmann equation (\ref{Friedmann}) can be reduced to
\begin{equation}\label{slFRW}
  H^2\simeq\frac{1}{3M_{p}^2}(\frac{1}{4}|K|\dot{\phi}^2).
\end{equation}
Then, according to the general assumption $\ddot{\phi}\ll H\dot{\phi}$ during the inflation \cite{Weinberg}, Eq.(\ref{EOM}) becomes
\begin{equation}\label{sleom}
6H(\dot{\phi}^2+K+r)\simeq -K_{\phi}\dot{\phi},
\end{equation}
where the dissipative rate $r$ parametrizes the dissipative strength of our model. When $r \ll 1$ (weak dissipation regime), dissipation is not strong enough to affect the background dynamical evolution of the inflaton field. However, the thermal fluctuations of the radiation energy density will modify the field fluctuations and affect the primordial spectrum of perturbations, which will be reflected in the results of the article. When $r \gg 1$ (strong dissipation regime), dissipation will dominate both the background dynamics and the fluctuations, which causes slow-roll conditions to be satisfied more easily.

Generally, the radiation energy in warm inflation does not vanish because vacuum energy is continuously being
dissipated at the rate $\dot{\rho}_{v}=-\Gamma \dot{\phi}^2$. So, we find $\Gamma \dot{\phi}^2$ in the right side of Eq.(\ref{radiation}) will act like a source term which is feeding in radiation energy, while the second term $4H\rho_{\gamma}$ in the left side of Eq.(\ref{radiation}) is a damping term that is depleting it away. When $H$, $\Gamma$, and $\phi$ are slowly varying, which is the so-called slow-roll approximation during inflation, there will be some nonzero steady-state point for $\rho_{\gamma}$, which means that radiation production can be considered quasi-stable \cite{Berera1995}, i.e. $\dot{\rho}_{\gamma} \simeq 0$. From Eq.(\ref{radiation}) we obtain the density of radiation
\begin{equation}\label{slradiation}
  \rho_{\gamma} \simeq\frac{3}{4}r\,\dot{\phi}^2,
\end{equation}
Using Eqs.(\ref{slFRW})$-$(\ref{slradiation}), a relation between $\rho_{\gamma}$ and $\rho_{\phi}$ is obtained
\begin{equation}\label{phirad}
  \rho_{\gamma}=\frac{r}{2(\mathcal{L}_{X}+r)}\epsilon \rho_{\phi},
\end{equation}
where $\mathcal{L}_{X}=\dot{\phi}^2+K$. The condition $\epsilon\ll1$ through stability analysis is rewritten in terms of Eq.(\ref{phirad}) as
\begin{equation}
  \rho_{\phi}\gg\frac{2(\mathcal{L}_{X}+r)}{r}\rho_{\gamma},
\end{equation}
which describes the epoch in which the warm $k$-inflation occurs. On the other hand, inflation ends when the Universe goes to the radiation-dominated rage, at a time when $\epsilon\simeq1$, which means $\rho_{\phi}\simeq\frac{2(\mathcal{L}_{X}+r)}{r}\rho_{\gamma}$. The number of $e$-folds at the end of inflation is given by
\begin{equation}\label{efold}
  N=\int_{t_i}^{t_e}Hdt=\int_{\phi_{i}}^{\phi_{e}}\frac{H}{\dot{\phi}}d\phi\simeq\frac{\sigma}{2\sqrt{3}M_{p}}\int_{\phi_{i}}^{\phi_{e}}\sqrt{-K(\phi)}d\phi,
\end{equation}
where $\phi_{i}$ is the initial value of the inflaton field and $\phi_{e}$ is the final value of the inflaton field.

\section{\label{sec3}scalar perturbations}
In this section, we will consider the scalar perturbations of the metric, since it is different from that of general potential-driven inflationary models or $k$-inflation in the cold inflationary scenario.  We consider the inhomogeneous scalar perturbations of the spatially flat FRW background described by the metric in the longitudinal gauge
\begin{equation}\label{pFRW}
  ds^2=(1+2\Phi)dt^2-a^2(t)(1-2\Psi)\delta_{ij}dx^{i}dx^{j},
\end{equation}
where $a(t)$ is the scale factor, $\Phi=\Phi(t,x)$ and $\Psi=\Psi(t,x)$ are the metric perturbations. $\Phi$ and $\Psi$ are gauge-invariant variables introduced by Bardeen \cite{Bardeen1980}. The dominant components of the Universe are constituted by the radiation and the inflaton field interacting through the dissipation term $\Gamma$. In momentum space, for the Fourier components $e^{ikx}$, with $k$ being the comoving wave number, the perturbed Einstein equations (we omit the subscript $k$) are
\begin{equation}\label{00}
  -3H(\dot{\Phi}+H\Phi)-\frac{k^2}{a^2}\Phi=\frac{1}{2M_{p}^2}\delta\rho,
\end{equation}
\begin{equation}\label{0i}
  \dot{\Phi}+H\Phi=\frac{1}{2M_{p}^2}(\rho+p)\delta u,
\end{equation}
\begin{equation}\label{ij}
  \ddot{\Phi}+4H\dot{\Phi}+(2\dot{H}+3H^2)\Phi=\frac{1}{2M_{p}^2}\delta p.
\end{equation}
In the above equations, the dot signifies a derivative with respect to $t$; $\rho$ and $p$ are the total energy density and pressure, respectively; $\delta\rho$ and $\delta p$ are the linear perturbations of the total energy density and pressure, respectively; $\delta u$ is the scalar part of the linear perturbation of the 4-velocity. And because the perturbation of the total energy-momentum tensor dose not give rise to the anisotropic stress, $\Psi=\Phi$ \cite{Weinberg}.

In warm inflation, the linear perturbations of the total energy density and pressure have the relation \cite{Mukhanov1992,Oliveria2001,Mukhanov2005}
\begin{equation}\label{3relation}
  \delta p=\frac{\partial p}{\partial \rho}\delta \rho +\frac{\partial p}{\partial S}\delta S \equiv c_s^2\delta \rho+\delta p^N,
\end{equation}
where $S$ denotes the entropy of system and $\delta p^{N}$ is the contribution to the perturbation of the pressure due to the variation of the effective equation of state that relates $p$ and $\rho$, or entropy perturbation. In terms of Eqs.(\ref{00}), (\ref{ij}), and (\ref{3relation}), we can obtain the evolution equation of comoving curvature perturbation $\mathcal{R}_{k}$ in the momentum space,
\begin{equation}\label{3Rt}
  \dot{\mathcal{R}}_k = - \frac{H}{\dot{H}} \left( \frac{\delta p^N }{2 M_p^2}  - \frac{c_s^2 k^2}{a^2} \Phi \right),
\end{equation}
where $\mathcal{R}_k \equiv \Phi + \frac{2 (\dot{\Phi} + H \Phi)}{3 H (1 + w)} (w=\frac{p}{\rho})$, and it is a gauge-invariant quantity \cite{Weinberg} that is a physical quantity to describe the scalar perturbations of the metric.

Next, we will consider how the entropy perturbation $\delta p^N$ influences the comoving curvature perturbation $\mathcal{R}_k$. First,  $\delta p^N$ can be expressed as the other perturbation quantities such as $\delta\rho, \delta\rho_{\gamma},\Phi$, and so on. According the conditions $\rho_{\gamma}\ll \rho_{\phi}$, $\dot{\rho}_{\gamma}\ll 4H\rho_{\gamma}$ and the perturbed Einstein equations (\ref{00})-(\ref{ij}), we find
\begin{eqnarray}\label{3entropy}
  \delta p^N &\simeq & - \frac{m_{p}^2}{4 \pi} (\dot{\Phi} + H \Phi) W - \frac{m_{p}^2}{4
  \pi a^2} \left( \frac{\dot{\phi}^2 + K}{3 \dot{\phi}^2 + K} - c_s^2 \right)
  k^2 \Phi \nonumber \\
   &\quad& -\frac{2 \rho_\gamma}{3 (3 \dot{\phi}^2 + K)} \left( K \frac{\delta
  \rho_\gamma}{\rho_\gamma} - \frac{2 K_{\phi} \dot{\phi}^3 a v}{(\dot{\phi}^2 + K) k}
  \right),
\end{eqnarray}
where $m_p^2=G^{-1}$, $v\equiv -\frac{k}{a}\delta u^{\gamma}$, and $W \simeq -\frac{\Gamma}{3\dot{\phi}^2+K}$. From Eq.(\ref{3entropy}), the entropy perturbations can be divided into two parts, i.e. $\delta p^N=S_{\Gamma}+S_{\Phi}$, and they are
\begin{eqnarray}
  S_{\Gamma} & =&  \frac{m_p^2}{4 \pi} (\dot{\Phi} + H \Phi) \frac{\Gamma}{3
  \dot{\phi}^2 + K} \nonumber \\
  &\quad& -\frac{2 \rho_\gamma}{3 (3 \dot{\phi}^2 + K)} \left( K \frac{\delta \rho_\gamma}{\rho_\gamma} - \frac{2 K_{\phi} \dot{\phi}^3 a
  v}{(\dot{\phi}^2 + K) k} \right), \label{3Gamma}  \\
  S_{\Phi} & =&  - \frac{m_p^2}{4 \pi a^2} \left( \frac{\dot{\phi}^2 + K}{3
  \dot{\phi}^2 + K} - c_s^2 \right) k^2 \Phi, \label{3Phi}
\end{eqnarray}
respectively. $S_\Gamma$ is induced by the dissipation coefficient $\Gamma$ and the radiation which play an important role in the entropy perturbations. While $S_\Phi$ is induced by the scalar perturbation of the metric $\Phi$. In the cold inflation, $\Gamma=0$, and $\rho_\gamma=0$; thus, $S_\Gamma=0$ and only the part $S_\Phi$ exist, which is consistent with the previous result \cite{Sri2009}.

Second, we calculate the perturbed equations of the inflaton field and radiation in the longitudinal gauge. During the warm inflation, the main components of the Universe are the inflaton field and radiation. Their total energy momentum is conservative, while each is not conservative. And each satisfies
\begin{eqnarray}
   T^{\mu(\gamma)}_{\phantom{\mu}\nu;\mu} & = & Q_{\nu}^{(\gamma)} \label{3Qr},  \\
   T^{\mu(\phi)}_{\phantom{\mu}\nu;\mu} & = & Q_{\nu}^{(\phi)}.  \label{3Qphi}
\end{eqnarray}
Because the total energy momentum is conservative, the interaction terms $Q_{\nu}^{(\gamma)}$ and $ Q_{\nu}^{(\phi)}$ satisfy the relation: $Q_{\nu}^{(\gamma)}+ Q_{\nu}^{(\phi)}=0$. From Ref.\cite{Mar2011}, we find the interaction terms can be expressed by
\begin{equation}\label{3interaction}
   Q_{\nu}^{(\phi)} = - Q_{\nu}^{(\gamma)} = - \Gamma u^{\mu(\phi)}\nabla_{\mu} \phi (x, t) \nabla_{\nu} \phi (x,t).
\end{equation}
Thus, the linear perturbation of interaction terms is
\begin{eqnarray}
  \delta Q_0^{(\gamma)} & = & - \delta Q_0^{(\phi)} = \delta \Gamma \dot{\phi}^2 +
  2 \Gamma \dot{\phi} \dot{\delta \phi} - \Gamma \dot{\phi}^2 \Phi,  \label{3Q0} \\
  \delta Q_i^{(\gamma)} & = & - \delta Q_i^{(\phi)} =  \Gamma\dot{\phi} \partial_i \delta \phi. \label{3Qi}
\end{eqnarray}
In terms of Eqs.(\ref{3Qr})-(\ref{3Qi}), we can get the perturbed equations of the inflaton field and radiation in the momentum space:
\begin{eqnarray} \label{3Phit}
  &&(3 \dot{\phi}^2 + K) \ddot{\delta \phi} +[(3 \dot{\phi}^2 +
  K)^{{\bf.}} + 3 H (3 \dot{\phi}^2 + K) + \Gamma] \dot{\delta \phi} \nonumber \\
  && +\left[ \left( K_{\phi} \ddot{\phi} + \frac{K_{\phi \phi}}{2} \dot{\phi}^2 +
  3 H K_{\phi} \dot{\phi} \right) + \frac{k^2}{a^2} (\dot{\phi}^2 + K) \right]
  \delta \phi \nonumber \\
   && =2 (3 \dot{\phi}^2 + 2 K) \dot{\phi} \dot{\Phi} + (6
  \ddot{\phi} \dot{\phi}^2 + 6 H \dot{\phi}^3 - \Gamma \dot{\phi}) \Phi -
  \delta \Gamma \dot{\phi},
\end{eqnarray}
\begin{equation}\label{3rhot}
   \dot{\delta \rho}_\gamma + 4 H \delta \rho_\gamma - \frac{4 k}{3 a}
  \rho_\gamma v - 4 \rho_\gamma \dot{\Phi} = \delta \Gamma \dot{\phi}^2 + 2 \Gamma
  \dot{\phi} \dot{\delta \phi} - \Gamma \dot{\phi}^2 \Phi,
\end{equation}
\begin{equation}\label{3v}
   \dot{v} + \frac{\Gamma \dot{\phi}^2}{\rho_r} v + \frac{k}{a}
  \left( \Phi + \frac{\delta \rho_\gamma}{4 \rho_\gamma} + \frac{3 \Gamma \dot{\phi}}{4
  \rho_\gamma} \delta \phi \right) = 0.
\end{equation}
Because we just care about the scalar perturbations on the large scale ($k\ll aH$), Eqs.(\ref{3Phit})-(\ref{3v}) can be approximately written as
\begin{eqnarray}
  \Phi & \simeq & \frac{4 \pi}{m_p^2 H} \left[ (\dot{\phi}^2 + K) \dot{\phi}
  \delta \phi - \frac{4}{3 k} \rho_\gamma a v \right],  \label{3phia}\\
  \frac{\delta \rho_\gamma}{\rho_\gamma} & \simeq & \frac{\delta \Gamma}{\Gamma} - \Phi,
  \label{3rhoa}\\
  v & \simeq & - \frac{k}{4 a H} \left( \Phi + \frac{\delta \rho_\gamma}{4 \rho_\gamma}
  + 3 H \frac{\delta \phi}{\dot{\phi}} \right), \label{3va}
\end{eqnarray}
where we have used the slow-roll conditions (\ref{slcon}). Furthermore, according to Eqs.(\ref{3Rt}) and (\ref{3phia})-(\ref{3va}), the evolution equation of the comoving curvature perturbation $\mathcal{R}_k$ on the large scale is
\begin{eqnarray}\label{3Rtlarge}
  \dot{\mathcal{R}}_k \simeq \frac{2 \Gamma \rho}{3 (3 \dot{\phi}^2 + K) (\rho + p)} \Phi .
\end{eqnarray}
From the above equation, the variation with time on the large scale of $\mathcal{R}_{k}$ will be influenced by the dissipation coefficient $\Gamma$. The result we obtain is similar to the previous result for the potential-driven warm inflation \cite{Oliveria2001}. When $\Gamma$ is very small or $r\ll1$ (weak dissipation regime), the source term in the above equation can be neglected and the comoving curvature perturbation $\mathcal{R}_{k}$ is adiabatic. The adiabatic perturbation will be mostly responsible for the primordial spectrum of scalar perturbations, which is consistent with the observable results \cite{Planck2015}. Thus in the following sections, we will only discuss in the weak dissipation regime so that $\dot{\mathcal{R}}_k \simeq 0$. In the following, we will show that although the thermal dissipation is weak, it can obviously affect the primordial spectrum of scalar perturbations.

\section{\label{sec4}thermal fluctuations of inflaton field}
In this section, we will consider the thermal fluctuations of the inflaton field in the spatially flat gauge which is defined to be the slicing in which there is no curvature. Since it is simple to calculate the thermal fluctuations in this gauge, and when the dissipation coefficient $\Gamma$ is independent of the temperature $T$, the evolution equations of thermal fluctuations and metric perturbations are decoupling, so it is easy to give the analytic results \cite{Moss2009}. Thus, in this section we will make use of the spatially flat gauge and assume that $\Gamma=\Gamma(\phi)$. The perturbed FRW metric in this gauge is
\begin{equation}\label{fFRW}
  d s^2 = (1 + 2 A) d t^2 - 2 a (t) \partial_i B d t d x^i - a^2 (t) \delta_{i
  j} d x^i d x^j,
\end{equation}
where $A(x,t)$ and $B(x,t)$ are scalar perturbations of the metric. Under the space-time background described by the metric (\ref{fFRW}), we expand the  inflaton field to linear order, $\phi(x,t)=\phi(t)+\delta \phi(x,t)$, where $\delta \phi(x,t)$ is the linear response due to the thermal stochastic noise $\xi(x,t)$ in thermal system \cite{Ramos1994,Hall2004}. Thus the perturbed equations of inflaton field (\ref{3Phit}) in the spatially flat gauge become
\begin{eqnarray}\label{4eom}
 (3 \dot{\phi}^2 + K) \ddot{\delta \phi}_k + [3 H (3 \dot{\phi}^2 &+& K) +
  \Gamma] \dot{\delta \phi}_k + \frac{k^2}{a^2} (\dot{\phi}^2 + K) \delta \phi_k \nonumber\\
   &=& (3 \dot{\phi}^2 + K) \dot{\phi} \dot{A} + (6 H \dot{\phi}^3 - \Gamma \dot{\phi}) A  \nonumber\\
   && + \frac{k^2}{a} (\dot{\phi}^2 + K^{}) \dot{\phi} B - \delta
  \Gamma \dot{\phi},
\end{eqnarray}
where we have used the slow-roll conditions (\ref{slcon}). And the perturbed Einstein equation becomes
\begin{eqnarray}
  3 H^2 A + \frac{k^2}{a} H B & = & - \frac{4 \pi}{m_p^2} \delta \rho, \label{4B}\\
  H A & = & \frac{4 \pi}{m_p^2} (\rho + p) \delta u. \label{4A}
\end{eqnarray}
The metric perturbations $A$ and $B$, which can be expressed by Eqs.(\ref{4B}) and (\ref{4A}), are substituted into Eq.(\ref{4eom}) to eliminate the metric perturbations in the perturbed inflaton equation. Thus Eq.(\ref{4eom}) becomes
\begin{equation}\label{4lang}
  \mathcal{L}_X c_s^{- 2} \ddot{\delta \phi}_k (t) + 3 H (\mathcal{L}_X c_s^{-
  2} + r) \dot{\delta \phi}_k (t) +\mathcal{L}_X \frac{k^2}{a^2} \delta \phi_k
  (t) = \xi_{k}(t),
\end{equation}
where we have introduced the thermal stochastic noise source $\xi_{k}(t)$ to truly describe the evolution of the thermal fluctuations. Equation (\ref{4lang}) is called the Langevin equation which is used to describe the behavior of a scalar field interacting with radiation \cite{Berera2000,Schwinger1961,Keldysh1964}. And in the above equation, we have used the relations $c_{s}^2=\frac{\dot p}{\dot \rho}\simeq \frac{\dot{\phi}^2+K}{3\dot{\phi}^2+K}$ and $\mathcal{L}_{X}=\dot{\phi}^2+K$. Because it is difficult to solve the Langevin equation analytically, we will consider a kind of  reasonable approximate approach to get a formal solution of the equation. Since we only care about the perturbation spectra when horizon crossing, which is well inside the slow-roll inflationary regime and the slow roll regime is over damped \cite{Liddle2000}, the inertia term $\ddot{\delta \phi}_k$ can be neglected in Eq.(\ref{4lang}). Thus, the Langevin equation becomes \cite{Berera2000,Taylor2000}
\begin{equation}\label{4langa}
  3 H (\mathcal{L}_X c_s^{- 2} + r) \dot{\delta \phi}_k (t) +\mathcal{L}_X
  \frac{k^2}{a^2} \delta \phi_k (t) = \xi_k(t).
\end{equation}
Thus the analytic formal solution of the above equation is
\begin{eqnarray}\label{4solution}
  \delta \phi_k (t) &=& \frac{1}{3 H (\mathcal{L}_X c_s^{- 2} + r)} \exp
  \left[ - \frac{t - t_0}{\tau (\phi_0)} \right] \times \nonumber \\
  && \times \int^t_{t_0} \exp \left[
  \frac{t' - t_0}{\tau (\phi_0)} \right] \xi_k (t') d t' \nonumber  \\
  && +\delta \phi_k (t_0)
  \exp \left[ - \frac{t - t_0}{\tau (\phi_0)} \right],
\end{eqnarray}
where $\tau (\phi) = \frac{3 H (\mathcal{L}_X c_s^{- 2} + r)}{\mathcal{L}_X
\frac{k^2}{a^2}} = \frac{3 H (\mathcal{L}_X c_s^{- 2} + r)}{\mathcal{L}_X
k_p^2}$ and $k_p$ is the physical wave number, which can be expressed as $k_p=k/a$. From the solution (\ref{4solution}), we find that the larger $k_{p}$ is, the faster the relaxation rate is. If $k_{p}$ of one mode of the perturbed inflaton field $\delta \phi_k$ is enough large to relax within a Hubble time, the mode will be thermal. Once the physical wave number $k_p$ to which one mode of $\delta \phi_k$ corresponds is smaller than the freeze-out physical wave number $k_F$, it will be not affected by thermal noise $\xi_k(t)$ during a Hubble time \cite{Berera2000}. Based on the above discussion, the freeze-out physical wave number $k_F$ from the formula $\frac{3 H (\mathcal{L}_X c_s^{- 2} + r)}{\mathcal{L}_X k_p^2}=1$ is worked out:
\begin{equation}\label{4freeze}
   k_F = \sqrt{\frac{3 H^2}{c_s^2} \left( 1 + \frac{r
  c_s^2}{\mathcal{L}_X} \right)}>H .
\end{equation}
From the above equation, the result in the warm $k$-inflation that the freezing time of thermal fluctuations of the inflaton field $\delta\phi_k$ is earlier than that of the horizon crossing is similar to that of the general potential-driven noncanonical warm inflationary models \cite{XiaoMin2014}. And when the noncanonical warm inflation model is reduced to the canonical warm inflation model, i.e., $c_s^2=1$ and $\mathcal{L}_{X}=1$, $k_F$ will be same as that of the canonical warm inflation model \cite{Taylor2000}. Generally, the thermal fluctuations of the inflaton field $\delta \phi_k$ in warm inflation can be expressed as $\delta \phi^2_{k}=\frac{k_F T}{2 \pi^2}$ \cite{Berera2000,Taylor2000}. Thus, from Eq.(\ref{4freeze}), we can obtain the thermal fluctuations of the inflaton field
\begin{equation}\label{4field}
   \delta \phi^2=\frac{H T}{2\pi^2}\sqrt{\frac{3}{c_{s}^2} \left( 1 + \frac{r c_{s}^2}{\mathcal{L}_X} \right)}.
\end{equation}
where we omit the subscript $k$. Since during the period of warm inflation $c_s<1$ and the temperature satisfies the relation $T>H$, the thermal fluctuations of the inflaton field are that $\delta \phi^2 > \frac{H^2}{4 \pi^2}$, while the vacuum fluctuations of the inflaton field are about $\frac{H^2}{4 \pi^2}$ \cite{Taylor2000}, so we find the density fluctuation in the warm $k$-inflation also mainly originates from the thermal fluctuations of the inflaton field, which is the fundamental characteristic of warm inflation perturbation theory.

\section{\label{sec5}perturbation spectra}
In this section, we will work out the spectra of scalar perturbations and tensor perturbations. First, the spectrum of scalar perturbations  $\mathcal{P}_{\mathcal{R}}$ and inflaton fluctuations $\delta\phi$  in cold inflation have the relation $\mathcal{P}_{\mathcal{R}} = \left( \frac{H}{\dot{\phi}} \right)^2 \delta \phi^2$. In warm inflation, when radiation production is quasistable and scalar perturbations are approximately adiabatic, the above relation can also estimate the power spectrum of scalar perturbations on the large scale \cite{Taylor2000}. According to the preceding analysis, warm $k$-inflation satisfies the above conditions. Thus we obtain the spectrum of scalar perturbations in the warm $k$-inflation:
\begin{eqnarray}\label{5scalar}
  \mathcal{P}_{\mathcal{R}_{warm}}&=&\frac{H^3 T}{2 \pi^2 \dot{\phi}^2} \sqrt{\frac{3}{c_s^2}
  \left( 1 + \frac{r c_s^2}{\mathcal{L}_X} \right)}  \nonumber \\
  & \simeq & \frac{2\sqrt{3}}{c_s} \frac{T}{H} \Bigg(\frac{H}{\dot{\phi}}\Bigg)^2 \Bigg(\frac{H}{2\pi}\Bigg)^2 \nonumber \\
  & = & \frac{2\sqrt{3}}{c_s} \frac{T}{H} \mathcal{P}_{\mathcal{R}_{cold}},
\end{eqnarray}
where we have used $\mathcal{P}_{\mathcal{R}_{cold}}=\Big(\frac{H}{\dot{\phi}}\Big)^2 \Big(\frac{H}{2\pi}\Big)^2$ and the condition in Eq.(\ref{slcon}):  $\frac{rc_{s}^2}{\mathcal{L}_{X}}\ll1-2c_{s}^2$. Since during warm inflationary period $T>H$ and $c_s<1$, from Eq.(\ref{5scalar}) we find the spectrum of scalar perturbations in our model is larger than that of general cold inflationary models, i.e.   $\mathcal{P}_{\mathcal{R}_{warm}}>\mathcal{P}_{\mathcal{R}_{cold}}$,
which will result in that the tensor-to-scalar ratio becomes small in the following discussion.
According to the formula $n_s - 1 = \frac{d \ln \mathcal{P}_{\mathcal{R}}}{d \ln k}$, we can obtain the spectral index of scalar perturbations:
\begin{equation}\label{5index}
  n_s - 1 \simeq  \left( \frac{1}{4} b - \frac{17}{4} \epsilon - \frac{3}{2} \eta- 3 \beta - \chi \right),
\end{equation}
where $\epsilon,\eta$, and $b$ are slow-roll parameters in Eq.(\ref{slcon}). They are all much less than unity within the inflationary period. While $\beta\equiv \frac{\ddot{\phi}}{H\dot{\phi}}$ and $\chi\equiv \frac{\dot{c_s}}{c_s H}$ in Eq.(\ref{5index}), they are also much less than unity during the inflationary period \cite{Weinberg,Mukhanov1999}. Therefore, the spectral index of scalar perturbations is nearly scale invariant, i.e., $n_s \simeq 1$, which is consistent with the observable result \cite{Planck2015}.

Next, we briefly discuss the spectrum of tensor perturbations. Since tensor perturbations do not couple to the thermal background and primordial gravitational waves are only generated by the quantum fluctuations of inflaton field as in cold inflation \cite{Taylor2000}, the spectrum and spectral index of tensor perturbations are given by usual expression
\begin{eqnarray}
  \mathcal{P}_{T_{warm}} &=& \frac{16 \pi}{m_p^2} \left( \frac{H}{\pi} \right)^2,  \label{5tensor}\\
    n_{T_{warm}} & = & - 2 \epsilon. \label{5tensorindex}
\end{eqnarray}
From Eqs.(\ref{5scalar}) and (\ref{5tensor}), we find the ratio of tensor to scalar is
\begin{equation}\label{5ratio}
  R_{warm}=\frac{\mathcal{P}_{T_{warm}} }{ \mathcal{P}_{\mathcal{R}_{warm}}}=\frac{\sqrt{3}}{6}\frac{H}{T}c_s R_{cold}=\frac{\sqrt{3}}{6}\frac{H}{T}16c_s\epsilon_{fz},
\end{equation}
where $R_{cold}$ is the ratio of the tensor to scalar in cold inflation, and $R_{cold}=\frac{\mathcal{P}_{\mathcal{R}_{cold}}}{\mathcal{P}_{T_{cold}}}=16\epsilon_{fz}$.
From Eq.(\ref{5ratio}), we find the ratio of the tensor to scalar is no longer a fixed value, which is so-called consistency relation \cite{Liddle1993} in cold inflation. Since $T>H$ and $c_s<1$ during the inflationary period, $R_{warm}<R_{cold}$. And in the original $k$-inflation \cite{Mukhanov}, $R=16c_s\epsilon_{fz}$ which is also lager than $R_{warm}$. Compared with some standard potential-driven inflationary models and the original $k$-inflation, $R_{warm}$ is more likely to be below the upper limit of $R$, and the observation predicts ($R<0.11$) \cite{Planck2015}. Therefore, warm $k$-inflation is different from general potential-driven cold inflationary models or $k$-inflation, and the analytic results computed by warm $k$-inflation may be easier to consist with the observable results.

\section{\label{sec6}specific cases}

To check  constraints on the model parameter $K(\phi)$ and the consistency of dissipation rate $r$ during the warm $k$-inflation by using the above results, we reconsider the specific cases in Ref.\cite{Peng2016}. In the reference, we have found that when $K(\phi)$ is a power-law function $-K_0^2\phi^{2n}$ or exponential function $-K_1^2 e^{2\alpha\phi}$, the dissipation coefficient $\Gamma$ can only take the function of $\phi$. Next, we will consider one of specific models because of the similarities of those models. Let us consider the example
\begin{equation}
 K(\phi)=-K_0^2\phi^{2n},
\end{equation}
where $K_0$ is a positive constant and $n$ is a positive integer. This form of $K(\phi)$ has been shown to be able to result in
kinetically driven inflation in Ref.\cite{Mukhanov1999}. And the dissipative coefficient $\Gamma$ takes $\Gamma_m \phi^{m}(m>4n)$; $m$ is an even integer. For simplicity, let $n=1$ and $m=6$, i.e., $K(\phi)=-K_0^2\phi^2$ and $\Gamma=\Gamma_6\phi^6$. Then, the slow-roll parameter $\epsilon$ can given by
\begin{equation}\label{epsilon}
\epsilon= \frac{4\sqrt{3}}{K_0 M_p} \left( \frac{M_p}{\phi}\right)^2,
\end{equation}
where $[K_0]=M_{p}^{-1}$. When the inflation ends, $\epsilon_{e}=1$. Thus the final value of the inflaton field is
\begin{equation}\label{endphi}
\left( \frac{\phi_e}{M_p} \right)^2 = \frac{4\sqrt{3}}{K_0 M_p}.
\end{equation}
 Moreover, the scalar perturbation (\ref{5scalar}) will cross outside the Hubble radius at approximately $50$ $e$-folds before the end of inflation based on the Planck 2015 data \cite{Planck2015}. In terms of Eq.(\ref{efold}),  we can get the following relation:
\begin{equation}\label{freeout}
\frac{K_0 M_p}{4\sqrt{3}}\left[ \left( \frac{\phi_{fz}}{M_p} \right)^2-\left( \frac{\phi_e}{M_p} \right)^2 \right] = 50.
\end{equation}
Then form Eqs.(\ref{epsilon})-(\ref{freeout}), we can obtain the value of $\epsilon$ at the freezing time of the scalar perturbation,
\begin{equation}\label{epsilonfz}
\frac{1}{\epsilon_{fz}}-1=50, ~ i.e.,~ \epsilon_{fz}=0.0196.
\end{equation}
From Eq.(\ref{5ratio}), we find
\begin{equation}
R_{warm}=\frac{\sqrt{3}}{6}\frac{H}{T}16c_s \epsilon_{fz} \simeq 0.091 c_s \frac{H}{T}.
\end{equation}
Because $c_s<1$ and $T>H$ during the warm $k$-inflation, the tensor-to-scalar ratio $R_{warm}<0.091$, which is below the upper limit of $R$ the Planck 2015 data give \cite{Planck2015}.

Finally, we will check the consistency of dissipation rate $r$. Because $\rho_{\gamma}=\frac{\pi^2 g_{\star}}{30} T^4 (g_{\star}\simeq 100)$, we can obtain the ratio $T/H$ based on Eqs. (\ref{slradiation}) and (\ref{5scalar}),
\begin{equation}
\frac{T}{H} \simeq \left( \frac{45}{4\pi^2 g_{\star}} \right)^{1/3} \left[ \frac{3}{c_s^2}\left(1+\frac{r c_s^2}{\mathcal{L}_{X}} \right) \right]^{1/3} \left( \frac{r}{\mathcal{P}_{\mathcal{R}_{warm}}}\right)^{1/3}.
\end{equation}
Because $\frac{r c_s^2}{\mathcal{L}_{X}} \ll 1$ and $c_s<1$ based on the consistency of warm $k$-inflation\cite{Peng2016}, the condition for warm inflation $T > H$ is approximately
\begin{equation}
r > g_{\star}\mathcal{P}_{\mathcal{R}_{warm}}.
\end{equation}
Since the cosmic microwave background observations give a scalar power spectrum of $10^{-10}$ on the large scales, very small amounts of the dissipation rate can contribute to warm $k$-inflation.

\section{\label{sec7}conclusions}

In this paper, we mainly investigate the theory of cosmology perturbations and perturbation spectra in a kind of novel warm inflationary model, so-called warm $k$-inflation. We first give the dynamical equations of homogenous inflaton field in the unperturbed FRW background. And the emerging condition and end condition of warm $k$-inflation are obtained. Second, since warm $k$-inflation is a kind of two-field inflationary model, one should consider entropy perturbations. Entropy perturbations provide a nontrivial evolution of the comoving curvature perturbation $\mathcal{R}$. Thus, we have discussed how the entropy perturbations in our model affect the evolution of the comoving curvature perturbation $\mathcal{R}$ in the longitudinal gauge. Then we find the entropy perturbations $\delta p^N$ could be divided into two parts, i.e., $ \delta p^N=S_{\Phi}+S_{\Gamma}$. The scalar perturbations of the metric $\Phi$ and $\Psi$ contribute to the part $S_{\Phi}$, and the dissipation coefficient $\Gamma$ plays a central role the part $S_{\Gamma}$ in the entropy perturbations. In addition, $\Gamma$ dominates the source term in the evolution equation of comoving curvature perturbation on the large scale. When the weak dissipation condition($r\ll 1$) is satisfied, the source term can be neglected such that the equation is reduced to $\dot{\mathcal{R}}\simeq 0$. As a result, the primordial spectrum of scalar perturbations is due only to adiabatic perturbations.

Furthermore, we find the thermal fluctuations of the inflaton field are larger than the vacuum fluctuations. Therefore we show that density fluctuations also mainly originate from the thermal fluctuations in the warm $k$-inflation, which is consistent with the basic feature of the cosmology perturbations in general warm inflationary models. Finally, the perturbation spectra and the tensor-to-scalar ratio are calculated. And we find the spectrum of scalar perturbations is larger than that in the general potential-driven cold inflationary models. On the contrary, the tensor-to-scalar ratio is smaller than the previous results. Further, considering a specific example, we find $R_{warm} < 0.091$, which is below the upper limit of the observable result, and show the weak dissipation rate can result in warm $k$-inflation. Thus our model is more likely to be consistent with the observable results.

A better understanding of entropy perturbations in the strong dissipation condition enables us to investigate how the entropy perturbations and comoving curvature perturbation influence each other. And when the dissipation coefficient $\Gamma$ is dependent of the temperature, which causes the equations of the inflaton field and radiation to be coupled,  we will solve the coupled equations with a numerical method.
\acknowledgments This work was supported by the National Natural Science Foundation of China (Grants No. 11575270,
No. 11175019, No. 11235003, and No. 11605100).

\end{document}